\documentclass[prl,twocolumn]{revtex4} 

\usepackage{graphics}

\begin{document}

\title{Low-field diffusion magneto-thermopower of a
 high mobility two-dimensional electron gas}

\author{S. Maximov, M. Gbordzoe,
 H. Buhmann and L.W. Molenkamp}

\affiliation{Physikalisches Institut (EP3), Universit\"at
W\"urzburg, Am Hubland, 97074 W\"urzburg, Germany}

\author{D. Reuter}
\affiliation{Lehrstuhl f\"ur Festk\"orperphysik,
Ruhr-Universit\"at Bochum, Universit\"atstr.\ 150, 44801 Bochum,
Germany}

\date{\today}

\begin{abstract}
The low magnetic field diffusion thermopower of a high mobility
GaAs-heterostructure has been measured directly on an
electrostatically defined micron-scale Hall-bar structure (4
$\mu$m $\times$ 8 $\mu$m) at low temperature ($T=1.6$~K) in the
low magnetic field regime ($B \le 1.2$~T) where delocalized
quantum Hall states do not influence the measurements. The sample
design allowed the determination of the field dependence of the
thermopower both parallel and perpendicular to the temperature
gradient, denoted respectively by $S_{xx}$ (longitudinal
thermopower) and $S_{yx}$ (Nernst-Ettinghausen coefficient). The
experimental data show clear oscillations in $S_{xx}$ and $S_{yx}$
due to the formation of Landau levels for $0.3$~T $<B\le 1.2$~T
and reveal that $S_{yx}\approx 120\, S_{xx}$ at a magnetic field
of 1~T, which agrees well with the theoretical prediction that the
ratio of these tensor components is dependent on the carrier
mobility: $S_{yx}/S_{xx}=2\omega_{c}\tau$.
\\



\end{abstract}

\maketitle

Thermopower experiments have been used extensively to obtain
information on transport and scattering in two-dimensional
electron gases (2DEGs) in compound semiconductors (for reviews,
see Refs.\ \cite{b.a} and \cite{b.b}). Because of the strong
electron phonon coupling in these systems, the experimental signal
is usually dominated by phonon-drag, hence, apart from the desired
electronic transport contributions, the signal also contains a
very significant contribution due to details of the
electron-phonon interaction. In order to extract the true
electronic or ``diffusion'' thermopower, usually drastic
approximations have to be made \cite{b.a,b.b}. It would thus be
very desirable to have an experimental approach that is not
influenced by phonon-drag effects and directly yields the
diffusion thermopower. In this paper we describe the development
of such an experiment.

We present direct measurements of the magnetic field dependence of
the diffusion thermopower using current heating techniques in
specially designed micro-Hall bar structures. The samples were
fabricated from high mobility GaAs-AlGaAs heterostructures [$\mu
\approx 100$~m$^2$/(Vs)] using split-gate techniques. A current
passing through an electron channel adjoining the Hall structure
is used to exclusively heat the electron gas, leaving the lattice
temperature unchanged. This current-heating technique has
previously been successfully used to determine the diffusion
thermopower of mesoscopic systems such as quantum point contacts
\cite {molenkampqpc} and quantum dots
\cite{staring,moeller,godijn}. The present sample design allows
the direct measurement of the tensor components of the thermopower
both parallel ($S_{xx}$) and perpendicular ($S_{yx}$) to the
temperature gradient in $x$-direction. The results are discussed
in the framework of theoretical models developed for the magnetic
field regime where the formation of Landau levels leads to a
modulation of the density of states \cite{b.f}, but does not yet
induce the formation of edge states. Therefore, the magnetic field
in the present study is restricted to the low field regime ($B\le
1.2$~T) where the influence of the quantum Hall effect can be
neglected.

Fig.~1 shows an SEM-photograph of the sample structure, including
a schematic diagram of the measurement. The micro-Hall bar and the
electron heating channel are defined by Schottky-gates, thus
forming the quantum point contacts (QPCs), which are used as
voltage probes. Gates A, D, E and F form the micro-Hall bar and
gates A, B, C and D the heating channel. Utilizing the fact that
the thermopower of a QPC is quantized \cite{molenkampqpc},
QPC$_{4}$ and QPC$_{5}$ are used to determine the electron
temperature in the channel $T_{ch}$ by measuring the voltage drop
$V_{25}\equiv V_5-V_2$ across the electron channel, while gates E
and F are not defined. We then have
$V_{25}=(S_{QPC5}-S_{QPC4})\Delta T_{ch}$, where $\Delta T_{ch}$
equals the temperature difference between the electrons in the
channel ($T_{ch}$) and in the surrounding 2DEG ($T_l\approx
1.6$~K), which is in thermal equilibrium with the crystal lattice:
$\Delta T_{ch}= T_{ch} - T_l$ \cite{b.7}. Note that the
temperature difference $\Delta T_{ch}$ enters here rather than a
gradient, since the thermovoltage across a QPC can only be
measured globally. Experimentally, one observes that $\Delta
T_{ch} \propto I^2$, where $I$ is the net heating current, as
expected from a simple heat balance equation that is valid for not
too large current values \cite{HvH}. Fig.~2 shows the
experimentally determined thermovoltage as a function of the
channel heating current. It can be seen that the parabolic
dependence is valid for currents up to 12 $\mu$A. For the
temperature calibration the thermopower of QPC$_4$ was adjusted to
$S_{QPC4}=20$~$\mu$V/K \cite{HvH} and the thermopower of QPC$_5$
was set at a minimal value ($S_{QPC5}\approx 0$). The temperature
calibration is given on the right axis of Fig.~2.


\begin{figure}[t]
    \begin{center}
        \resizebox{7cm}{8.24cm}{\includegraphics{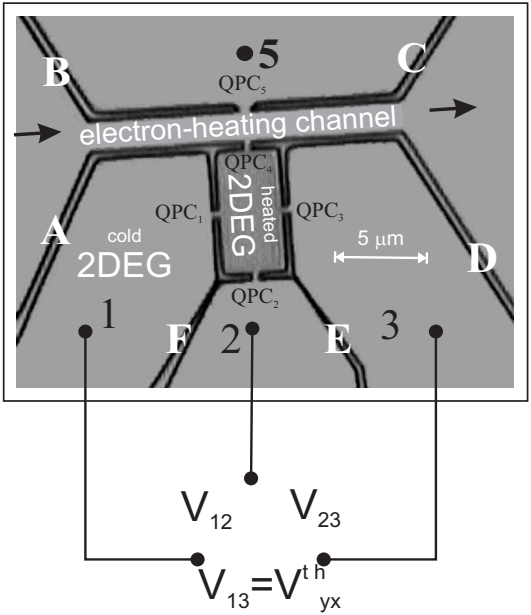}}
    \end{center}
    \caption{SEM-photograph of the split-gates structure and the
scheme of the measurements.}
\end{figure}


For a thermopower experiment on the micro-Hall bar, QPC$_{4}$ was
adjusted into the tunneling regime ($G_{QPC4} \approx 3\times
10^{-5}$ $\Omega^{-1}$ $< e^2/h$). QPC$_{1}$, QPC$_{2}$ and
QPC$_{3}$ were set to higher conductance values ($G_{QPC}\approx
10\times2 e^2/h$) in order to keep their thermopower minimal
($S_{QPC1,2,3}\approx 0$). The channel current was set to
$\sim$10~$\mu$A which yields an electron temperature in the
channel of $T_{ch} \approx 6.6$~K [c.f.\ Fig.~2]; this current
value gave a good compromise between pronounced thermovoltage
signals and the avoidance of lattice heating effects. The inset of
Fig.~2 shows the longitudinal resistance of the channel at this
current level of 10~$\mu$A; evidently the Shubnikov-de Haas
oscillations are nearly suppressed, which ensures an approximately
constant heat dissipation over the field range studied.

The experiments were carried out at a temperature of about 1.6~K
in an $^4$He cryostat equipped with a 10~T superconducting magnet.
The 2DEG carrier density ($2.8\times10^{15}$ m$^{-2}$) and
mobility ($\approx 100$ m$^{2}$(Vs)$^{-1}$) were obtained from
Hall and Shubnikov-de Haas (SdH) measurements. Standard lock-in
amplifier measurement techniques were used to measure the
thermoelectric effects. As mentioned above, the 2DEG heating is
proportional to $I^2$. Using ac-currents ($I=i_0\cos(\omega t)$)
and a lock-in detection of the second harmonic ($2\omega$), the
measured signal solely depends on the thermovoltage
[$V_{th}\propto I^2 =\left(i_0\cos(\omega t)\right)^2\propto
i_0^2\cos(2\omega t)$].


\begin{figure}[t]
    \begin{center}
        \resizebox{7.5cm}{5.4cm}{\includegraphics{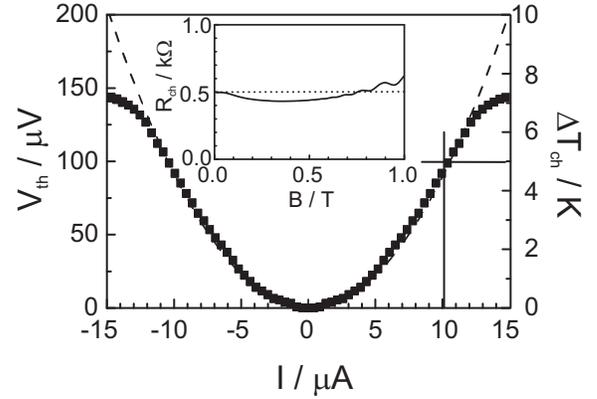}}
    \end{center}
    \caption{Electron temperature as a function of the channel heating
current. The solid line with squares represents the measured data;
the dotted line is a parabolic fit. Inset: Suppression of the SdH
oscillations in the channel at a heating current of 10~$\mu$A. The
difference between the dotted line and the minimum of the SdH is
about 15\%.}
\end{figure}


Fig.~1 indicates how the two tensor components of the thermopower
can be obtained in our current heating experiment. First, we note
that the thermopower of a 2 DEG in a magnetic field is a local
property, so that the thermovoltages we measure are proportional
to a temperature gradient, $V_{th} = S \nabla_x T dx$. We assume
that the temperature gradient across the micro-Hall bar coincides
with the line 4-2 connecting QPC$_4$ and QPC$_2$ defining the
$x$-direction. An important parameter for the experiment is the
electron temperature at the crossing of line 4-2 and the line
connecting QPC$_1$ and QPC$_3$ (line 1-3 defining the
$y$-direction). If the electron temperature outside the micro-Hall
bar is assumed to be equal to the lattice temperature, a
temperature gradient is expected to develop between the side which
is in contact with the heating channel ($T_e^{max}\approx T_{ch}$)
and the surrounding 2DEG ($T_l$). In zero order approximation a
constant temperature gradient along the line connecting QPC$_2$
and QPC$_4$ would have the following form:
$\nabla_xT_e=(T_e^{max}-T_l)/x_0\approx 5$~K$/8$~$\mu$m $=
0.625$~K/$\mu$m, where $x_0$ is the extension of the micro-Hall
bar in $x$-direction ($x_0=8$~$\mu$m).

From Fig.~1, it is clear that $V^{th}_{yx}$, the thermovoltage
perpendicular to the temperature gradient, can be determined by
measuring the voltage difference between the areas 1 and 3,
$V^{th}_{yx}\equiv V_{13}\equiv V_3-V_1$, provided the intrinsic
thermopower of QPC$_1$ and QPC$_3$ can be neglected. For
$V^{th}_{xx}$ however, the required voltage probe at the crossing
point of the lines 1-3 and 4-2 is not available. Instead, we can
obtain $V^{th}_{xx}$ from measuring the signals present at
$V_{12}\equiv V_2 - V_1$ and $V_{32}\equiv V_2 - V_3$. Since
$V_{12}$ and $V_{23}$ contain contributions from $V^{th}_{xx}$ as
well as $V_{yx}^{th}$, $V_{xx}^{th}$ can be determined by adding
$V_{12}$ and $V_{23}$ and subtracting $V_{13}\equiv V^{th}_{yx}$.
This allows us to compare $V_{xx}^{th}$ and $V_{yx}^{th}$ directly
without an exact knowledge of $\nabla_x T_e$ in the micro-Hall
structure.


\begin{figure}[t]
    \begin{center}
        \resizebox{7.5cm}{5.6cm}{\includegraphics{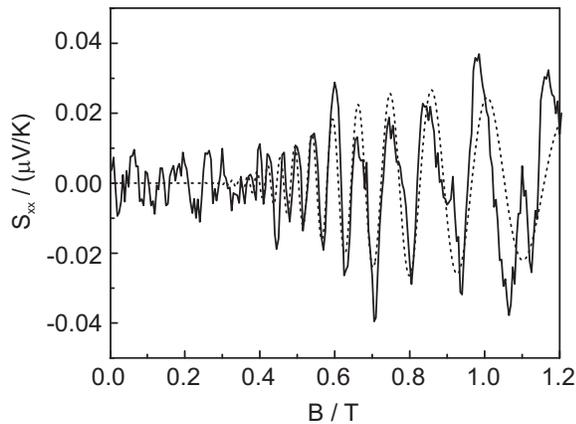}}
    \end{center}
    \caption{Thermopower $S_{xx}$ parallel to the temperature
gradient: The solid line corresponds to the experimental data for
$\Delta T=$2.5~K and the dashed dotted line represents the
calculation according to Eq.~3.}
\end{figure}


For both components, the experiments show clear oscillations in
the thermopower with magnetic field (Figs.~3 and 4). The
thermopower signal has been calculated from the thermovoltage
measurement assuming the linear temperature gradient as mentioned
above. The presented field range studied can be separated into two
parts: First, $B < 0.3$~T; the electrons are considered as
classical particles which are deflected by the applied magnetic
field and scattered elastically at the device boundaries (mean
free path $l_{mfp}\approx 8$~$\mu$m) \cite{ballistic}, and second,
$0.3 < B < 1.2$~T where the oscillations correspond to the
formation of Landau levels in the 2DEG and hence to the magnetic
field dependent modulation of the density of states. In the
following, we will present a detailed quantitative discussion of
the second magnetic field regime ($0.3 < B < 1.2$~T).

According to Ref.~\cite{b.f} the magnetic field behaviour of the
thermopower oscillations can, in the regime of Landau level
formation, be described by the following equations:

\begin{eqnarray}
S_{xx}&=&\frac{2}{1+\omega_{c}^2\tau^2}\left(\frac{\pi
k_B}{e}\right)D'(X) \nonumber \\
&&\times \exp{\left(-\frac{2\pi^2 k_B
T_D}{\hbar\omega_c}\right)}\sin{\left(\frac{2\pi
f}{B}-\pi\right)}\label{e.1}\\
\nonumber \\
S_{yx}&=&\frac{4\omega_c\tau}{1+\omega_{c}^2\tau^2}\left(\frac{\pi
k_B}{e}\right)D'(X)  \nonumber \\
&&\times \exp{\left(-\frac{2\pi^2 k_B
T_D}{\hbar\omega_c}\right)}\sin{\left(\frac{2\pi
f}{B}-\pi\right)}\label{e.2}
\end{eqnarray}

where $T_D$ is the Dingle temperature, $\omega_c$ the cyclotron
frequency, $\tau$ the transport relaxation time, and $f$ is the
frequency of the oscillations ($f/B=E_F/\hbar\omega_c$, where
$E_F$ is the Fermi energy). The quantity $D'(X)$ is the derivative
of the thermal damping factor $D(X)$, defined by $D(X)=X/\sinh{X}$
where $X=2\pi^2k_{B}T/\hbar\omega_c$. These equations were
originally \cite{b.f} derived for conditions where
$\omega_c\tau<1$, which would restrict the validity in our case to
magnetic fields up to $B\sim 20$ mT. However, Coleridge \emph{et
al.}~\cite{b.g} have shown that Eqs.~(3) and (4) are valid up to
much higher field values ($B \sim 1$~T) when localization effects
can be neglected (as it is the case for high mobility 2DEGs).


\begin{figure}[t]
    \begin{center}
        \resizebox{7.5cm}{5.6cm}{\includegraphics{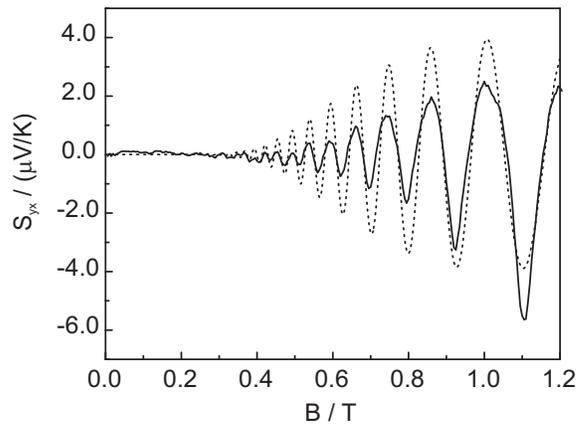}}
    \end{center}
    \caption{Thermopower $S_{yx}$ perpendicular to the temperature
gradient: The solid line corresponds to the experimental data for
$\Delta T=$2.5~K and the dashed dotted line represents the
calculation according to Eq.~4.}
\end{figure}


The results of the measurements up to 1.2~T are presented in
Fig.~3 and Fig.~4 together with fits using Eqns.~(\ref{e.1}) and
(\ref{e.2}). For the fits, the carrier density $n$ and the
mobility $\mu$ were taken from the transport characterization. The
Dingle temperature was obtained from the assumption that the
quantum mobility is approximately 10 times lower than the electron
mobility i.e. $T_D\approx 10\pi/\mu\approx 0.4$~K \cite{b.f,b.g}.
The thermal smearing was fitted by a free parameter
$\overline{T}_e$, which can be interpreted as the average electron
temperature in the micro-Hall bar. The best fits for an average
electron temperature of $\overline{T}_e=4$~K is in very good
agreement with the estimates made above concerning the temperature
gradient.

Both $S_{xx}$ and $S_{yx}$ can be fitted satisfactorily using the
same set of parameters, even though the amplitudes are very
different. According to Eqns.~(3) and (4), the ratio of the
thermopower perpendicular and parallel to the temperature gradient
is given by $S_{yx}/S_{xx} = 2 \omega_c\tau$. For the present
sample, the measured ratio at $B= 1$ T is $\approx 120$. Again,
this value agrees well with the expected value of 160. To our
knowledge this is the first successful measurement of the
diffusion thermopower for a semiconductor 2DEG system. The current
heating approach allows us to avoid the influence of phonon-drag
effects \cite{b.f,b.10,b.11}. From the consistency of the average
temperatures and the temperature gradients, which are obtained
from the fitting and the channel temperature calibration, it can
be concluded that the chosen geometry and the measurement
configuration are suitable for investigating the diffusion
thermopower of high mobility samples.
\\

Summarizing, the results presented here demonstrate that electron
heating techniques can be used to measure directly the
longitudinal and transverse components of the diffusion
thermopower. For low magnetic fields, thermopower fluctuations are
observed which originate from quasi-ballistic electrons; for
higher fields, the modulation of the electron density of states
due to Landau level formation determines the oscillatory part of
the diffusion thermopower. Current theories~\cite{b.b,b.f}
describe this oscillatory behavior to a large extent and can be
used to independently gauge the electron temperature. The
consistency of these measurements with theory opens up the way for
an alternative method for studying the diffusion thermopower in
the QHE as well as in the fractional quantum Hall effect regime,
where currently, experimental data are discussed controversially
\cite{b.a,b.12,b.13,b.14}.


This work performed with financial support from the Deutsche
Forschungsgemeinschaft (DFG Mo 771/5-2).


\begin{thebibliography}{99}
\bibitem{ballistic} For a qualitative and quantitative discussion
of the low magnetic field regime ($0\le B < 0.3$~T) a detailed
theoretical modelling of the ballistic thermopower of this special
sample geometry is needed including a numerical calculation of the
related transmission probabilities \protect{\cite{Pogosov}}, which
will be given in a later publication.
\end{thebibliography}
\end{document}